\documentclass{aa}  
\usepackage{graphicx}
\usepackage{txfonts}
\usepackage{graphicx}
\usepackage{lscape}
\usepackage{bm}
\usepackage[colorlinks=true, linkcolor=blue, citecolor=blue, filecolor=blue, urlcolor=blue]{hyperref}
\usepackage{graphicx}
\usepackage{txfonts}
\usepackage[version=3]{mhchem}
\usepackage{xcolor}
\usepackage{lscape}
\usepackage{ulem}
\usepackage{mhchem}

\usepackage{booktabs}
\usepackage{hyperref}

\usepackage[figuresright]{rotating}
\usepackage{mathtools}
\usepackage{float}
\usepackage{csquotes}

\begin{document} 

   \title{The SOFIA Massive (SOMA) Star Formation Q-band follow-up. II.}
   
   \subtitle{Hydrogen recombination lines toward high-mass protostars}
   
    \author{Prasanta Gorai\inst{1,2}\fnmsep\thanks{prasanta.astro@gmail.com}, Kotomi Taniguchi\inst{3}, Jonathan C. Tan \inst{4,5}, Miguel G$\acute{\rm {o}}$mez-Garrido\inst{6},
    Viviana Rosero\inst{7}, Izaskun Jim\'enez-Serra\inst{8}, Yichen Zhang\inst{9}, Giuliana Cosentino\inst{10}, Chi-Yan Law\inst{11}, Rub{\'e}n Fedriani\inst{12}, Gemma Busquet\inst{13,14,15}, Brandt A. L. Gaches\inst{16},  Maryam Saberi\inst{1,2}, Ankan Das\inst{17}
          }

   \institute{Rosseland Centre for Solar Physics, University of Oslo, PO Box 1029 Blindern, 0315, Oslo, Norway
              \and
              Institute of Theoretical Astrophysics, University of Oslo, PO Box 1029 Blindern, 0315, Oslo, Norway
              \email{prasanta.astro@gmail.com}
              \and
              National Astronomical Observatory of Japan, National Institutes of Natural Sciences,
              2-21-1 Osawa, Mitaka, Tokyo, 181-8588, Japan
              \and
              Department of Space, Earth \& Environment, Chalmers University of Technology, 412 93  Gothenburg, Sweden
              \and
              Department of Astronomy, University of Virginia, Charlottesville, VA 22904, USA
              \and
              Observatorio Astronomico Nacional (OAN-IGN), Alfonso XII 3, 28014, Madrid, Apain
              \and
              Division of Physics, Mathematics, and Astronomy, California Institute of Technology, Pasadena, CA 91125, USA
              \and 
              Centro de Astrobiologia (CAB), CSIC-INTA, Ctra. de Ajalvir km 4, Torrejon de
              Ardoz, E-28806, Spain
              \and
              Department of Astronomy, Shanghai Jiao Tong University, 800 Dongchuan Rd., Minhang, Shanghai 200240, People’s Republic of China
             \and
              Institut de Radioastronomie Millimétrique, 300 Rue de la Piscine, 38400 Saint-Martin-d'Hères, France
              \and
              Osservatorio Astrofisico di Arcetri, Largo Enrico Fermi, 5, 50125 Firenze FI, Italy
              \and
              Instituto de Astrof\'isica de Andaluc\'ia, CSIC, Glorieta de la Astronom\'ia s/n, E-18008 Granada, Spain
              \and
              Departament de F\'{\i}sica Qu\`antica i Astrof\'{\i}sica (FQA), Universitat de Barcelona (UB), Mart\'{i} i Franqu\`es 1, 08028 Barcelona, Spain
              \and
              Institut de Ci\`encies del Cosmos (ICCUB), Universitat de Barcelona (UB), Mart\'{i} i Franqu\`es 1, 08028 Barcelona, Spain
              \and
              Institut d\'Estudis Espacials de Catalunya (IEEC), Esteve Terradas 1, edifici RDIT, Parc Mediterrani de la Tecnologia (PMT) Campus del Baix Llobregat—UPC
              \and
              Faculty of Physics, University of Duisburg-Essen, Lotharstraße 1, 47057 Duisburg, Germany
              \and
              Institute of Astronomy Space and Earth Sciences, P 177, CIT Road, Scheme 7m, Kolkata 700054, West Bengal, India
              }


\abstract
{Hydrogen recombination lines (HRLs) are valuable diagnostics of the physical conditions in ionized regions surrounding high-mass stars. Understanding these lines, including broadening mechanisms and intensity trends, can provide insights into HII region densities, temperatures, and kinematics.}
{This study aims to investigate the physical properties of ionized gas around massive protostars by analyzing the hydrogen recombination lines (H$\alpha$ and H$\beta$) in the Q-band.}
{We carried out observations using the Yebes 40m radio telescope in the Q-band (30.5–50 GHz) toward six high-mass protostars selected from the SOMA Survey (G45.12+0.13, G45.47+0.05, G28.20-0.05, G35.20-0.74, G19.08-0.29, and G31.28+0.06). The observed line profiles were analyzed to assess broadening mechanisms, and electron densities and temperatures were derived. The results were compared with available Q-band data from the TianMa 65-m Radio Telescope (TMRT), as reported in the literature, and ALMA Band 1 (35–50 GHz) Science Verification observations toward Orion KL, analyzed in this study.}
{A total of eight H$\alpha$ (n = 51 to 58) and ten H$\beta$ (n = 64 to 73) lines were detected toward G45.12+0.13, G45.47+0.05, and G28.20-0.05 and non detection in other sources. We derive electron densities of $\rm{\sim 1-5\times 10^6\ cm^{-3}}$ and temperatures of 8000–10000~K for the sources. However, for Orion KL, we obtained one order of magnitude lower electron density, while its temperature is found to be more similar. Interestingly, G45.12 and G28.20 show an increasing intensity trend with frequency for both H$\alpha$ and H$\beta$ transitions, contrary to the decreasing trend observed in Orion KL.}
{The line widths of the detected HRLs indicate contributions from both thermal and dynamical broadening, suggesting the presence of high-temperature ionized gas that is likely kinematically broadened (e.g., due to turbulence, outflows, rapid rotation, or stellar winds). Pressure broadening caused by electron density may also have a minor effect. We discuss different scenarios to explain the measured line widths of the HRLs. The contrasting intensity trends between the sources may reflect variations in local physical conditions or radiative transfer effects, highlighting the need for further investigation through higher-resolution observations and detailed modeling.} 
    \keywords{stars: massive – stars: formation – HII regions – ISM: general --Ionization}
    \authorrunning{Gorai, Taniguchi, Tan, et al.}
  \maketitle

\section{Introduction}
High-mass protostars and recently formed massive stars
may be identified by the presence of small HII regions, known as hyper-compact (HC) or ultra-compact (UC) HII regions if their radii are less than 0.01 and 0.1 parsecs, respectively \citep[e.g.,][]{wood1989,kurtz1994,gaume95,depree2004,giveon2005,keto08a,churchell2010}. These HII regions are expected to be created either by protostellar outflow shock ionization, especially in the earliest protostellar phases \citep[e.g.,][]{2024ApJ...967..145G}, or by photo-ionization by Extreme ultraviolet (EUV) radiation emitted from massive stars, including during the later stages of the main protostellar accretion phase \citep[e.g.,][]{2016ApJ...818...52T,2017ApJ...849..133T} \citep[see, e.g.,][for a review]{2014prpl.conf..149T}. A general evolutionary sequence of HII region expansion is expected as gas density in the vicinity of protostars decreases.
 
For example, electron densities of a sample of HC HII regions have been found to be $\gtrsim
10^5$ cm$^{-3}$, at least an order of magnitude greater than those of UC HII regions with $\gtrsim
10^4$ cm$^{-3}$ \citep[][]{kurtz2005}. Radio recombination lines (RRLs) are high principal quantum $n$ spectral lines emitted by electrons recombining with a positive ion. These lines, occurring in the radio frequency portion of the electromagnetic spectrum, are commonly observed in astrophysical environments, including HC and UC HII regions \citep{churchwell1990}. Radio recombination lines have been widely used in astrophysics, especially as probes of the physical conditions of the plasma \citep[e.g.,][]{alves2015}.

Previous surveys of radio recombination lines have primarily utilized single-dish telescopes operating at centimeter (cm) wavelengths. These cm-RRLs typically exhibit principal quantum numbers $n\geq 66$ and have been observed with angular resolutions normally of about a few arc-minutes. Examples of such surveys include those conducted by \citet{lockman1989,caswell1987,anderson2009,anderson2014,alves2015}. Furthermore, several studies at higher resolutions, using interferometers that focus on individual sources \citep[e.g.,][]{gaume95,sewi04,depree2004,Sewilo2008,keto08b,zhang19}, have provided valuable insights into the characteristics of HC and UC HII regions. For example, these investigations have uncovered significantly broadened line widths within these regions, which tend to diminish as the HII region size increases. Hydrogen recombination lines (HRLs) have also provided key information to constrain the small-scale physical structure in HC HII regions, revealing the presence of ionized disks, winds, and jets toward Cepheus A HW2 and MonR2-IRS2 \citep{Jimenez-serra2011,Jimenez-serra2013,Jimenez-serra2020}.

Millimeter and (sub)millimeter HRLs toward HII regions in a large sample of clumps from the APEX Telescope Large Area Survey of the Galaxy (ATLASGAL) have been reported using single-dish telescopes, including the IRAM 30m, Mopra 22m, and APEX 12m \citep{kim2017,kim2018}. \cite{liu2022} conducted a Q-band survey toward the Orion KL using the Tianma 65 m radio telescope (TMRT) and detected 177 RRLs. Among these, 126 were hydrogen RRLs, 40 were helium RRLs, and 11 were carbon RRLs, with maximum changes in principal quantum number ($\Delta$n) of 16, 7, and 3, respectively. Their result suggests hydrogen and helium RRLs arise from M42 while carbon RRLs originate from the photodissociation region (PDR). Following this, \cite{liu2023} also reported the detection of RRLs of ions heavier than helium using the TMRT telescope's multi-band (12-50 GHz) line survey of Orion KL.

Here, we present Q-band (30.5-50 GHz) line survey data towards six massive protostars. We report the detection of 18 RRLs towards G45.12+0.13 (hereafter G45.12), G45.47+0.05 (hereafter G45.47), and G28.20-0.05 (hereafter G28.20); nondetection in three sources, G35.20-0.74 (hereafter G35.20), G19.08-0.29 (hereafter G19.08), and G31.28+0.06 (hereafter G31.28).  Line analysis of RRLs provides an estimation of electron densities and temperatures. We discuss the possible mechanisms of line broadening, including thermal, dynamical, and pressure contributions.  In addition, we employ a theoretical analysis to constrain the possible electron temperatures and densities of the HII regions associated with the protostars and compare these estimates with the observed results.

This paper is organized as follows: Section \ref{sec:2} describes observational details and data analysis procedures. Results are presented in Section \ref{sec:3}. Discussion is presented in Section \ref{sec:4}, and finally, in Section \ref{sec:5}, we provide concluding remarks.

\section{Observations, data reduction, and targets \label{sec:2}}

\begin{table*}
\caption{Summary of target sources}
\centering{
\begin{tabular}{ccccccccc}
\hline
Source& R.A. & Dec.& $d$ & $d_G$& $v_{\rm {LSR}}$& $L_{\rm bol}$ & $EM$& S$_\nu$\\
(Type) & (J2000) &(J2000) &(kpc) & (kpc)& (km/s) & ($L_{\odot}$)$^{(d)}$ &(pc cm$^{-6}$)&(Jy)\\
\hline
G28.20-0.05 (I) & $18^{\rm {h}}42^{\rm {m}}58\fs12$ &  $-4\degr13\arcmin57\farcs644$ & 5.7 &4.1& 95.6 & $1.4^{+1.9}_{-0.95} \times 10^{5}$ &$\rm{3.6\times10^8}$(a)&0.630(e)\\
G31.28+0.06 (I) & $18^{\rm {h}}48^{\rm {m}}11\fs82$ &$-01\degr26\arcmin31\farcs01$&4.9&4.97&110.0&$1.03 \times 10^{5}$&&0.234(f)\\
G45.47+0.05 (II) & $19^{\rm {h}}14^{\rm {m}}25\fs74$ &  $+11\degr09\arcmin25\farcs90$ & 8.4 &5.80& 61.5 & $3.5^{+2.4}_{-1.4} \times10^{5}$&$\rm{4.8\times10^7}$(b)&0.200(g)\\
G19.08-0.29	(II) &$18^{\rm {h}}26^{\rm {m}}48\fs43$ &$-12\degr26\arcmin28\farcs04$&4.5&3.84&65.4&&&--\\
G35.20-0.74 (III) & $18^{\rm {h}}58^{\rm {m}}13\fs03$ &  +$01\degr40\arcmin36\farcs14$ & 2.2 &6.45& 32.0 & $6.6^{+3.7}_{-2.4} \times 10^{4}$& $\rm{3.3\times10^8}$ (c)&0.010(h)\\
G45.12+0.13 (III) & $19^{\rm {h}}13^{\rm {m}}27\fs96$ &  $+10\degr53\arcmin35\farcs690$ & 7.4 &5.77& 59.5 & $8.0^{+6.5}_{-3.6} \times10^{5}$ & $\rm{1.5\times10^9}$(b)& 5.400(g)\\

\hline
\end{tabular}}
\tablebib{$d$: distance; $d_G$: Galactocentric distance; $EM$: emission measure; 
a) \cite{Sewilo2008}; b) \cite{wood1989}; (c) \cite{Beltran2016}; (d) \cite{Telkamp2025,law22}; \cite{Minier2005}, (e) \cite{Sewilo2011}, (f) \cite{codella2010}, (g) \cite{hofner1999}, (h) \cite{Beltran2016}- total flux obtained by adding flux from all 1.3 centimeter continuum sources as provided in their Table 2 and Fig. 1b, S$_\nu$ is the flux density obtained from Very Large Array (VLA) 1.3 cm observations.}
\label{tab:sources}
\end{table*}

\begin{figure*}
\centering
\includegraphics[width=\textwidth]{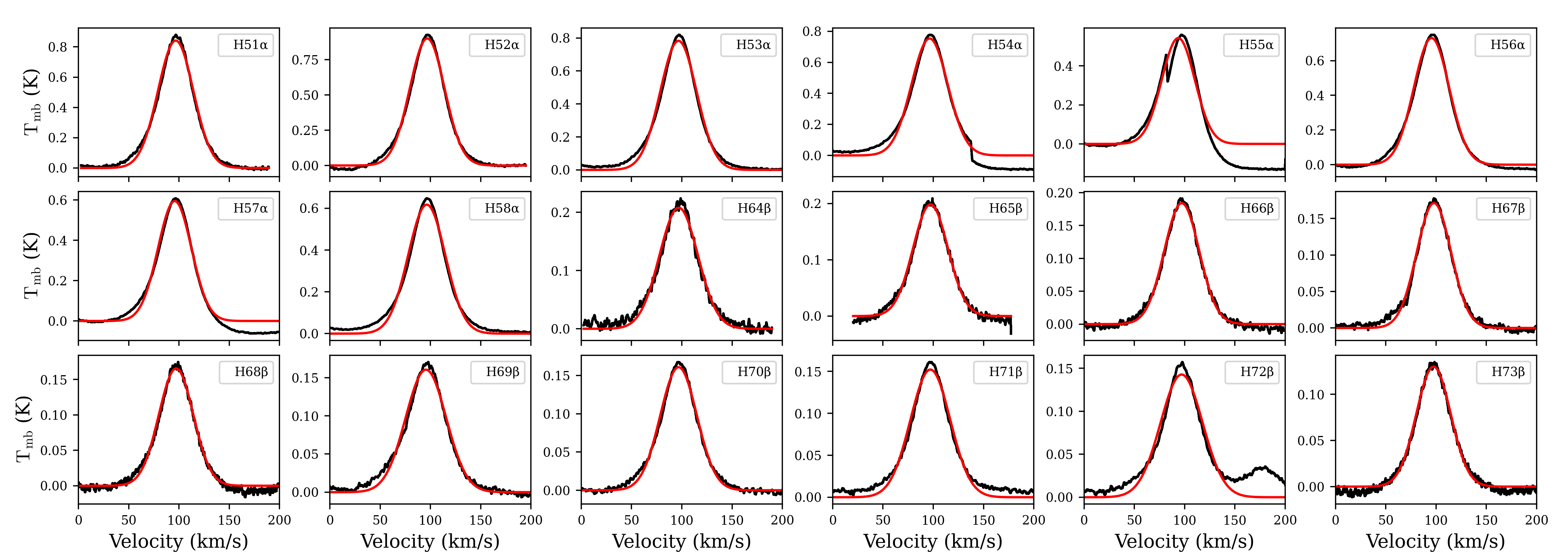}
\caption{The observed and Gaussian fitted spectra of H$\alpha$ and H$\beta$ lines towards G45.12+0.13. The black line represents observed H$\alpha$ and H$\beta$ lines toward G45.12+0.13, and the red line depicts the Gaussian fitted spectra.}
\label{fig:recom1}
\end{figure*}

\begin{figure*}
\centering
\includegraphics[width=\textwidth]{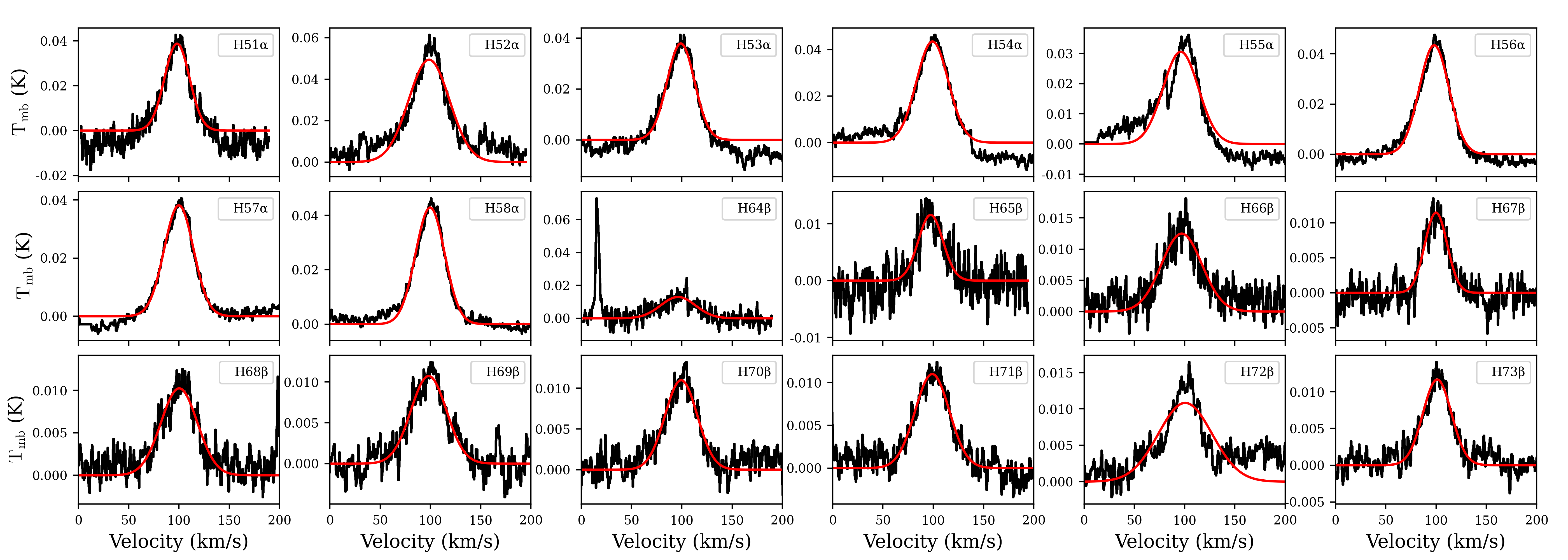}
\caption{The observed and Gaussian fitted spectra of H$\alpha$ and H$\beta$ lines towards G45.47+0.05. The black line represents observed H$\alpha$ and H$\beta$ lines toward G45.47+0.05, and the red line depicts the Gaussian fitted spectra.}
\label{fig:recom2}
\end{figure*}

\begin{figure*}
\centering
\includegraphics[width=\textwidth]{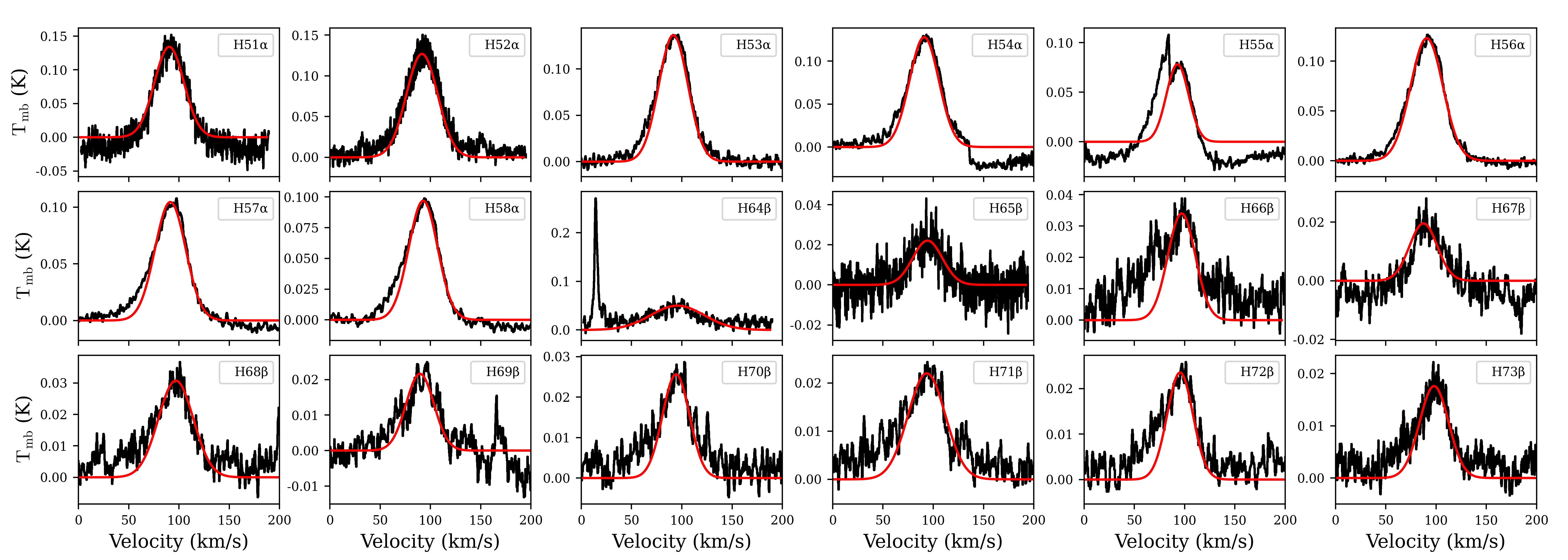}
\caption{The observed and Gaussian fitted spectra of H$\alpha$ and H$\beta$ lines towards  G28.20-0.05. The black line represents observed H$\alpha$ and H$\beta$ lines toward G28.20-0.05, and the red line depicts the Gaussian fitted spectra.}
\label{fig:recom3}
\end{figure*}

\subsection{Observations and data reduction}
The line survey data presented in this paper are part of two Yebes 40m radio telescope (RT40m) Q band proposals (Proposal IDs - 21A004 and 23A017; PI - Prasanta Gorai), which cover six high-mass protostars that are part of the SOFIA Massive (SOMA) Star Formation survey (PI - J. Tan) \citep{DeBuizer2017,Liu2019,Liu2020,Fedriani2023,Telkamp2025}. The coordinates of our targets (G28.20, G45.47, G45.12, G35.20, G19.08, G31.28), along with their line of sight velocities ($v_{\rm{LSR}}$) and distances, are summarized in Table \ref{tab:sources}. 

For the source G28.20, observations were carried out using the Yebes 40m single-dish radio telescope from 25th to 27th October 2021 with a total observing time of 15 hours. The other five sources were observed from 30th March to 18th April 2023, with a total observation time of 40 hours. The survey was carried out using new receivers built within the Nanocosmos project \citep{Tercero2021}, which consisted of two cooled high electron mobility transistor (HEMT) amplifiers covering the 31.52-49.97~GHz band with dual polarizations. Fast Fourier transform spectrometers (FFTSs) with 8$\times$2.5 GHz bandwidth with a spectral resolution of 38.15~kHz provide full coverage of the Q-band in both polarizations. These observations were performed using the standard position-switching mode. The beam size varies from $\sim$ 54$^{''}$  at 32 GHz to 36$^{''}$ at 48 GHz. The calibration was performed at the beginning of the position-switching, observing the sky and both hot and cold loads, and repeating this procedure every 18 min. Pointing and focus were corrected hourly through pseudo-continuum observations of intense SiO maser lines towards evolved stars close to our target sources. The pointing errors were within 7$^{"}$, and the calibration uncertainties are estimated to be less than 15$\%$.

The off-source positions were regions where the visual extinction ($A_V$) is below three mag in the $A_V$ maps obtained from the Atras and Catalogue of Dark Clouds \citep{dobashi2005}\footnote{\url{https://darkclouds.u-gakugei.ac.jp/more/readme.html}}. For our analysis, we converted the observed antenna temperature ($T_{A}$) to main beam brightness temperature ($T_{\rm mb}$ using the main beam efficiency, $B_{\rm eff}$ = 0.738\ exp(-($\nu$(GHz)/72.2)$^{2}$), and forward efficiency, $F_{\rm eff}$ =0.97 \citep{Tercero2021}, using the formula $T_{\rm mb}$ = $T_{A}\times(F_{\rm eff}/B_{\rm eff}$). We used the CLASS software from the GILDAS\footnote{\url{https://www.iram.fr/IRAMFR/GILDAS/}} package to create FITS files by reducing and combining data from the different dates of observations. We conducted line analyses in the CASSIS software \citep{vast15}. We have detected 18 recombination lines towards three sources: G28.20, G45.12, and G45.47. In contrast, there is no RRL identification towards the other three sources: G35.20, G19.08, and G31.28. Hence, in the remainder of the paper, we will focus solely on the targets where we have detected HRLs.

\subsection{Targets}
The sources are selected from the SOFIA Massive (SOMA) star formation survey \citep{DeBuizer2017,Liu2019,Liu2020,Fedriani2023,Telkamp2025}, which used SOFIA-FORCAST to measure 8 to 40~$\rm \mu$m emission of a sample $\gtrsim$ 50 sources. The SOMA sources have well-measured IR spectral energy distributions (SEDs), thus allowing detailed constraints on the physical properties of the sources.  Our six target sources belong to three different groups of SOMA sample: Type I: ``MIR sources in IRDCs'' - relatively isolated sources in Infrared Dark Clouds; Type II:``Hypercompact''- often jet-like, radio sources, where the MIR emission extends beyond the observed radio emission; Type III: ``Ultra-compact'' - radio sources where the radio emission is more extended than the MIR emission. The luminosity of all these sources is within the range $10^{4}-10^{5}\:L_\odot$. In our sample, we have two sources of each type. The source coordinate (R.A., Dec.), distance, systematic velocity ($v_{\rm LSR}$), and bolometric luminosity are provided in Table \ref{tab:sources}. In the following sections, brief descriptions of G45.12, G45.47, and G28.20 sources are provided.

\subsubsection{G45.12+00.13}
G45.12+00.13 (IRAS 19111+1048) has a measured far kinematic distance of 7.4~kpc \citep{ginsburg2011}. The radio morphology of this region shows a highly inhomogeneous ionized medium \citep{vig2006}, which is consistent with the extended mid-infrared (MIR) morphology \citep{Liu2019}. \cite{vig2006} proposed that the source is an embedded cluster of zero-age main sequence (ZAMS) stars with 20 compact sources, including one nonthermal source, identified by their radio emission. The central UCHII source S14 is deduced to be of spectral type O6 from the integrated radio emission. However, recent results from \cite{Sequeira2025}, as part of the SOMA Radio survey, only detected the very compact point source S20 (i.e., the nonthermal source) and the innermost part of the UC HII region source S14. They concluded that most of the emission to the northwest of the central UC HII region arises from an extended and diffuse irregular cloud of ionized material, rather than a cluster of compact sources. Based on SOFIA images, \citet{Liu2019} found MIR to far-infrared (FIR) emission peaking at the S14 position. They did not find a distinct source at the position of G45.12+0.13 west, and concluded that the MIR extension to the southwest of S14 could be due to blueshifted outflows, which are also revealed in near-infrared (NIR). Previously, several HRLs such as H30$\alpha$, H76$\alpha$, H99$\alpha$, H110$\alpha$ were reported in this source \citep{wood1989,arya2002,churchell2010,Tan2020}.

\subsubsection{G45.47+00.05}
G45.47+0.05 was first detected as a UC HII region in the radio continuum at 6 cm \citep{wood1989} and lies at a distance of 8.4~kpc. A bipolar wide-angle ionized outflow was discovered from the massive protostar G45.47+0.05 using VLA and ALMA observations \citep{rosero2019,Zhang2019}. The H30$\alpha$ recombination line showed strong maser amplification. It also suggested that there is a photoevaporation flow launched from a disk of radius 110 au, with an electron temperature ($T_e$) of 10$^{4}$\ K and an electron number density ($n_e$) of $\rm{1.5\times10^7\ cm^{-3}}$ \citep{Zhang2019}. High angular resolution (0.4$^{''}$) VLA data revealed G45.47 is resolved into two centimeter continuum sources separated by 0.4$^{''}$ \citep{rosero2019}. \citep{zhang19} also reported the presence of an ionized jet candidate seen toward the south of the central, brightest component, possibly of nonthermal nature.

\subsubsection{G28.20-0.05} 
G28.20-0.05 is an isolated massive protostar associated with a HC HII region and located at a distance of 5.7 kpc \citep{Sewilo2008}. There is compelling evidence that ionizing feedback occurs within its protostellar core. This feedback manifests through the ionization of its disk wind and, notably, some surrounding denser gas structures. This is corroborated by a ring emitting centimeter to millimeter free-free radiation \citep{Sewilo2008,Sewilo2011,law22}. This source shows chemically rich properties along with recombination lines \citep{Gorai2024,Law2025}.

\begin{table*}[ht]
\scriptsize{
\centering
\caption{Summary of Gaussian fit of detected recombination lines towards three high-mass sources \label{tab:lines}}
\begin{tabular}{|c|c|ccc|ccc|ccc|}
\hline
Line &Freq. (MHz)
& \multicolumn{3}{c|}{G45.12+0.13}
& \multicolumn{3}{c|}{G45.47+0.05}
& \multicolumn{3}{c|}{G28.20--0.05} \\
\cline{3-11}
& & Peak (mK) & FWHM (km/s) & Int. (K\,km/s)
& Peak (mK) & FWHM (km/s) & Int. (K\,km/s)
& Peak (mK) & FWHM (km/s) & Int. (K\,km/s) \\
\hline
H51$\alpha$ & 48153.60 & 841.65 $\pm$ 2.18 & 41.46 $\pm$ 0.12 & 36.99 $\pm$ 0.15 & 38.74 $\pm$ 0.71 & 29.30 $\pm$ 0.62 & 1.20 $\pm$ 0.03 & 133.99 $\pm$ 1.23 & 34.60 $\pm$ 1.89 & 4.91 $\pm$ 0.27 \\
H52$\alpha$ & 45453.72 & 900.77 $\pm$ 2.15 & 40.78 $\pm$ 0.11 & 38.93 $\pm$ 0.14 & 49.39 $\pm$ 0.61 & 47.83 $\pm$ 0.69 & 2.50 $\pm$ 0.05 & 126.91 $\pm$ 1.07 & 36.68 $\pm$ 1.92 & 4.93 $\pm$ 0.26 \\
H53$\alpha$ & 42951.97 & 782.54 $\pm$ 2.72 & 42.66 $\pm$ 0.17 & 35.39 $\pm$ 0.19 & 37.86 $\pm$ 0.49 & 32.21 $\pm$ 0.48 & 1.29 $\pm$ 0.03 & 136.15 $\pm$ 0.43 & 34.24 $\pm$ 0.60 & 4.94 $\pm$ 0.09 \\
H54$\alpha$ & 40630.50 & 753.57 $\pm$ 6.28 & 42.51 $\pm$ 0.41 & 33.95 $\pm$ 0.43 & 43.43 $\pm$ 0.56 & 37.37 $\pm$ 0.56 & 1.72 $\pm$ 0.03 & 127.79 $\pm$ 0.43 & 34.36 $\pm$ 0.65 & 4.65 $\pm$ 0.09 \\
H55$\alpha$ & 38473.36 & 542.20 $\pm$ 8.92 & 38.65 $\pm$ 0.73 & 22.21 $\pm$ 0.56 & 30.57 $\pm$ 0.55 & 41.48 $\pm$ 0.86 & 1.34 $\pm$ 0.04& 78.17 $\pm$ 0.49 & 26.09 $\pm$ 0.53 & 2.16 $\pm$ 0.05 \\
H56$\alpha$ & 36466.26 & 730.18 $\pm$ 2.54 & 41.55 $\pm$ 0.17 & 32.16 $\pm$ 0.17 & 43.41 $\pm$ 0.33 & 34.78 $\pm$ 0.31 & 1.60 $\pm$ 0.02 & 123.03 $\pm$ 0.44 & 37.97 $\pm$ 0.96 & 4.95 $\pm$ 0.13 \\
H57$\alpha$ & 34596.38 & 594.39 $\pm$ 4.44 & 39.22 $\pm$ 0.34 & 24.71 $\pm$ 0.28 & 38.16 $\pm$ 0.28 & 34.30 $\pm$ 0.29 & 1.39 $\pm$ 0.02 & 104.53 $\pm$ 0.57 & 35.88 $\pm$ 1.19 & 3.98 $\pm$ 0.13 \\
H58$\alpha$ & 32852.20 & 617.31 $\pm$ 3.62 & 43.68 $\pm$ 0.30 & 28.58 $\pm$ 0.26 & 42.86 $\pm$ 0.31 & 34.07 $\pm$ 0.28 & 1.55 $\pm$ 0.02 & 96.86 $\pm$ 0.27 & 33.93 $\pm$ 0.50 & 3.48 $\pm$ 0.05 \\
H64$\beta$ & 47914.18 & 207.91 $\pm$ 0.89 & 45.10 $\pm$ 0.22 & 9.94 $\pm$ 0.06 & 12.97 $\pm$ 1.01 & 40.74 $\pm$ 3.68 & 0.56 $\pm$ 0.07 & --& -- &-- \\
H65$\beta$ & 45768.44 & 198.44 $\pm$ 0.68 & 40.97 $\pm$ 0.16 & 8.62 $\pm$ 0.04 & 11.56 $\pm$ 0.37 & 28.90 $\pm$ 1.07 & 0.36 $\pm$ 0.02 & 22.06 $\pm$ 0.66 & 34.68 $\pm$ 5.53 & 0.81 $\pm$ 0.13 \\
H66$\beta$ & 43748.95 & 184.08 $\pm$ 0.54 & 40.96 $\pm$ 0.14 & 7.99 $\pm$ 0.04 & 12.49 $\pm$ 0.24 & 47.38 $\pm$ 1.05 & 0.63 $\pm$ 0.02 & 33.99 $\pm$ 0.58 & 32.17 $\pm$ 2.70 & 1.16 $\pm$ 0.10 \\
H67$\beta$ & 41846.55 & 171.17 $\pm$ 0.58 & 40.26 $\pm$ 0.16 & 7.31 $\pm$ 0.04 & 11.45 $\pm$ 0.24 & 26.17 $\pm$ 0.65 & 0.32 $\pm$ 0.01  & 19.55 $\pm$ 0.62 & 32.51 $\pm$ 5.30 & 0.67 $\pm$ 0.11 \\
H68$\beta$ & 40052.88 & 165.47 $\pm$ 0.45 & 40.38 $\pm$ 0.13 & 7.08 $\pm$ 0.03 & 10.24 $\pm$ 0.28 & 42.68 $\pm$ 1.34 & 0.46 $\pm$ 0.02  & 30.75 $\pm$ 0.49 & 40.87 $\pm$ 4.98 & 1.33 $\pm$ 0.16 \\
H69$\beta$ & 38360.27 & 160.42 $\pm$ 0.77 & 47.59 $\pm$ 0.26 & 8.09 $\pm$ 0.06 & 10.73 $\pm$ 0.18 & 41.69 $\pm$ 0.80 & 0.47 $\pm$ 0.01  & 21.65 $\pm$ 0.51 & 32.60 $\pm$ 4.22 & 0.75 $\pm$ 0.10 \\
H70$\beta$ & 36761.72 & 160.76 $\pm$ 0.64 & 43.02 $\pm$ 0.20 & 7.33 $\pm$ 0.04 & 10.98 $\pm$ 0.19 & 37.86 $\pm$ 0.75 & 0.44 $\pm$ 0.01 & 25.62 $\pm$ 0.41 & 31.58 $\pm$ 2.31 & 0.86 $\pm$ 0.06 \\
H71$\beta$ & 35250.77 & 152.08 $\pm$ 0.98 & 46.71 $\pm$ 0.35 & 7.53 $\pm$ 0.07 & 11.00 $\pm$ 0.20 & 40.26 $\pm$ 0.86 & 0.47 $\pm$ 0.01 & 22.01 $\pm$ 0.31 & 43.66 $\pm$ 5.05 & 1.02 $\pm$ 0.12 \\
H72$\beta$ & 33821.51 & 142.69 $\pm$ 1.69 & 49.32 $\pm$ 0.67 & 7.46 $\pm$ 0.13 & --& -- & --  & 23.50 $\pm$ 0.32 & 30.24 $\pm$ 1.77 & 0.75 $\pm$ 0.05 \\
H73$\beta$ & 32468.48 & 130.03 $\pm$ 0.37 & 40.33 $\pm$ 0.13 & 5.56 $\pm$ 0.02 & 11.67 $\pm$ 0.19 & 33.40 $\pm$ 0.64 & 0.41 $\pm$ 0.01 & 17.60 $\pm$ 0.28 & 34.06 $\pm$ 3.19 & 0.64 $\pm$ 0.06 \\
\hline
\end{tabular}
}
\end{table*}

\section{Results \label{sec:3}}

\subsection{Identification of RRLs and observed line parameters \label{sec:3.1}}

The rest frequency of RRLs can be expressed as
\begin{equation} \label{eq_rrlfreq}
\nu^{\rm RRL}_{\rm rest} = R_m c
\left( \frac{1}{n^2}-\frac{1}{(n+\Delta n)^2} \right),
\end{equation} 
where 
\begin{equation} 
R_m = \frac{R_\infty}{(1+(m_e/m_n))}.
\end{equation}
Here, $c$ is the  speed of light, $R_\infty$ is the Rydberg constant ($R_\infty=
109737.31568$ cm$^{-1}$), $m_e$
is the mass of the electron, and $m_n$ is the mass of the corresponding neutral atom. We have six sources in our sample (see Table 1). We have detected  8  H$\alpha$ transitions and 10 H$\beta$ transitions towards three sources. None of these transitions are detected towards G35.20, G19.08, and G31.28, and we will not discuss these sources in the following sections. Figures \ref{fig:recom1}, \ref{fig:recom2}, and \ref{fig:recom3} depict the observed recombination lines toward G45.12, G45.47, and G28.20, respectively. Table \ref{tab:lines} summarizes all the detected transitions.

\subsection{Properties of millimeter hydrogen recombination lines \label{sec:3.2}}
Figures \ref{fig:recom1}, \ref{fig:recom2}, and \ref{fig:recom3} show the observed and Gaussian fitted spectra towards G45.12, G45.47 and G28.20, respectively. The line width (FWHM, $\Delta v$) and peak intensity ($T_{\rm mb}$) of each observed HRL are measured by fitting a single Gaussian to the observed spectrum.  However, the fitting result obtained for the H55$\alpha$ line may not be accurate due to the discontinuity feature in the observed profile. The discontinuity in the observed profile is due to the emission being detected in two different sections of the FFTS. The line parameters of all the observed transitions, such as rest frequency ($\nu_0$), full-width at half maximum (FWHM), intensity, and integrated intensity, are summarized in Table \ref{tab:lines}.  
We find that the line width of G45.12 is around 40 km s$^{-1}$, $\sim$ 30-40 km s$^{-1}$ in G45.47 and $\sim$ 30-35 km s$^{-1}$ for G28.20 (see \ref{tab:lines}). If we consider a temperature $T_e= 10^{4}\:$K, the thermal line width of hydrogen is 21.4 km s$^{-1}$. Hence, thermal broadening alone cannot explain the observed line width, which requires nonthermal motion of the gas contributing to the broadening. 

Table \ref{tab:fwhm-ratio} shows the linewidth ratios of detected H$n\alpha$ (i.e., $n =$ 51, 55, 53, 54, 55, 57, 58) to the H56$\alpha$ and  H$n\beta$ ($n =$ 64, 65, 66, 67, 68, 69, 79, 70 72, 73) to the H71$\beta$.  The ratios are close to unity for all transitions, which indicates that different transitions are probing the same gas under the same physical conditions. Some relative differences may be due to microturbulence and different gas kinematics, such as rotation and outflow.  The line width of the recombination lines is discussed in more detail in Section \ref{sec:3.3}. 

\begin{figure*}
\centering
\includegraphics[width=\textwidth]{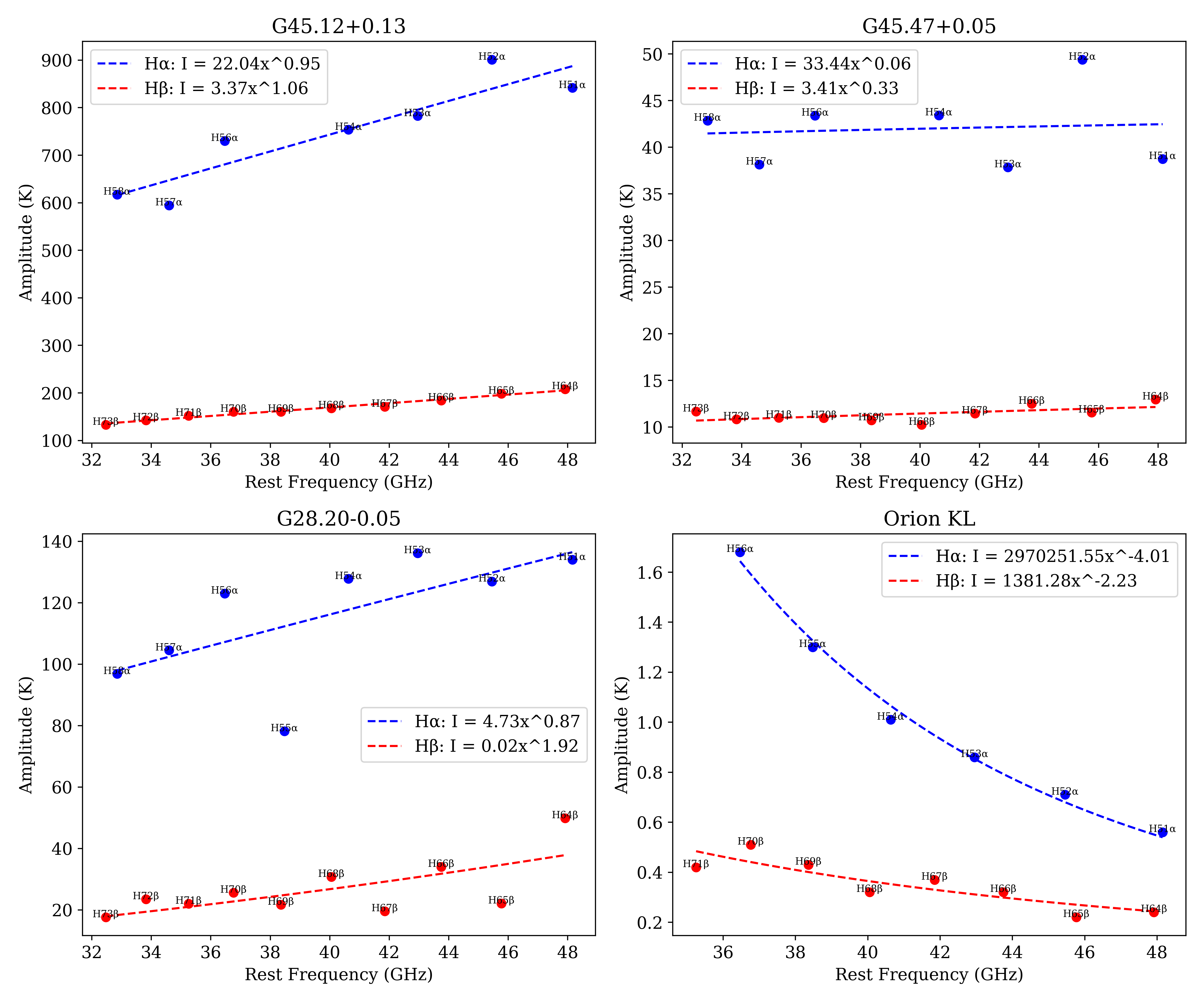}
\caption{Comparison of the observed intensity of H$\alpha$ and H$\beta$ towards our target sources and Orion KL.}
\label{fig:int-variation}
\end{figure*}

Figure \ref{fig:int-variation} shows the observed intensity variation with frequency for the H$\alpha$ and H$\beta$ lines. We apply a power-law fit to the observed trend of HRL intensity with frequency. For G45.12 and G28.20, both H$\alpha$ and H$\beta$ lines exhibit positive spectral indices (slopes close to or above 1), suggesting a nearly linear or mildly increasing trend of intensity with frequency. In G45.47, the H$\alpha$ line shows an almost flat dependence (slope $\sim$ 0), while H$\beta$ has a slightly rising trend, indicating a weaker frequency dependence. Furthermore, we compared our results with single-dish observations of Orion KL \citep{liu2022}. In contrast to our target sources, Orion KL shows steep negative slopes for both H$\alpha$ (-4.01) and H$\beta$ (-2.23), indicating that the line intensities decrease sharply with frequency.

Comparing peak intensity ratios of mm-RRLs is useful for diagnosing whether they are emitted under local thermodynamic equilibrium (LTE) or non-LTE conditions. The LTE line ratio between different transitions can be estimated as follows \citep{Dupree1970}:
\begin{equation}
   R_{\rm LTE} =n^{2}{f_{n,n+\Delta n}} / ({m^{2} f_{m,m+\delta m}})
\end{equation}
Where $f$ is the oscillator strength, $n$ is the lower level of a $n\alpha$ line, and $m$ is the lower level of a higher-order line (here H$\beta$) at nearly the same frequency. The oscillator strength values have been taken from \cite{Menzel1968}. 
Observed line ratio and LTE ratio are provided in Table \ref{tab:lte-ratio}. The peak-intensity ratios of both H$\alpha$ and H$\beta$ lines are, on average, consistent with the corresponding LTE ratios. Furthermore, we included the results of Orion KL in the same table and found that the estimated ratios slightly deviate from the LTE value.

\begin{table}[hbt]
  \centering
  \caption{Ratios of Hn$\alpha$ Line width to  H56$\alpha$ and ratios of Hn$\beta$ Line width to H71$\beta$  }
  \begin{tabular}{cccccc}
  \hline
    {Ratio} & G45.12& G45.47 &G28.20 &Orion KL$^{*}$ \\
    \hline
    H51$\alpha$/H56$\alpha$  &1.00  &0.84 &0.91&0.93 \\
    H52$\alpha$/H56$\alpha$  &0.98  &1.38 &0.97&0.95 \\
    H53$\alpha$/H56$\alpha$  & 1.03 &0.93 &0.90& 0.96\\
    H54$\alpha$/H56$\alpha$  &1.02  &1.07 &0.90& 0.94\\
    H55$\alpha$/H56$\alpha$  &0.93  &1.19 &0.69&0.94\\
    H57$\alpha$/H56$\alpha$  & 0.94 &0.99 &0.94&-- \\
    H58$\alpha$/H56$\alpha$  & 1.05 &0.98 &0.89&-- \\
    H64$\beta$/H71$\beta$    &0.97  &1.01 &--&0.74\\
    H65$\beta$/H71$\beta$    &0.88  &0.72 &0.79&0.95\\
    H66$\beta$/H71$\beta$    &0.88  &1.18 &0.74&1.02\\
    H67$\beta$/H71$\beta$    &0.86  &0.65 &0.74&1.02\\
    H68$\beta$/H71$\beta$    &0.86  &1.06 &0.94&1.00\\
    H69$\beta$/H71$\beta$    &1.02  &1.04 &0.75&1.00\\
    H70$\beta$/H71$\beta$    & 0.92 &0.94 &0.72&0.99\\
    H72$\beta$/H71$\beta$    &1.06  &-- &0.69&1.05\\
    H73$\beta$/H71$\beta$    &0.86  &0.83 &0.78&--\\
    \hline
  \end{tabular}
  \label{tab:fwhm-ratio}
  Note:  $^{*}$ The line width values are from \cite{liu2022}.
\end{table}

\begin{table}[hbt]
\caption{Comparison of LTE-predicted and observed line intensity ratios}
\begin{tabular}{|c|p{1cm}|c|c|c|c|}
\hline
Ratio & LTE Value & \multicolumn{4}{|c|}{Observed Ratio} \\
&&G45.47&G45.12&G28.20&Orion KL$^{a}$\\
\hline
$ \frac{H64\beta}{H51\alpha} $ & 0.278 &0.25&0.25&0.28&0.43\\
&&&&&\\
$ \frac{H65\beta}{H52\alpha} $ & 0.274 &0.20&0.22&0.25&0.31\\
&&&&&\\
$ \frac{H67\beta}{H53\alpha} $ & 0.283 &0.25&0.22&0.19&0.43\\
&&&&&\\
$ \frac{H68\beta}{H54\alpha} $ & 0.280 &0.25&0.23&0.27&0.32\\
&&&&&\\
$ \frac{H69\beta}{H55\alpha} $ & 0.277 &0.33&0.30&0.25&0.33\\
&&&&&\\
$ \frac{H70\beta}{H56\alpha} $ & 0.274 &0.25&0.22&0.23&0.30\\
&&&&&\\
$ \frac{H72\beta}{H57\alpha} $ & 0.282 &0.25&0.24&0.27&--\\
&&&&&\\
$ \frac{H73\beta}{H58\alpha} $ & 0.279 &0.25&0.22&0.27&--\\
\hline
\end{tabular}
\label{tab:lte-ratio}
Note- $^{a}$Data taken from \cite{liu2022}
\end{table}

\subsection{Line width of recombination lines \label{sec:3.3}}

As described in \cite{madrid2012}, several mechanisms may cause the broadening of RRL line widths. These include: (a) thermal/microturbulence Gaussian broadening, caused by the motion of emitting particles and gas parcels on small scales; (b) pressure (Stark) broadening due to the perturbation of atomic energy levels by the electric field from nearby charged particles; (c) dynamical broadening arising from bulk gas flows, such as infall, rotation and outflow. A comprehensive discussion of these processes and the underlying physics of RLs is given by \cite{gordon2002}.

\begin{figure*}
\centering
\includegraphics[width=\textwidth]{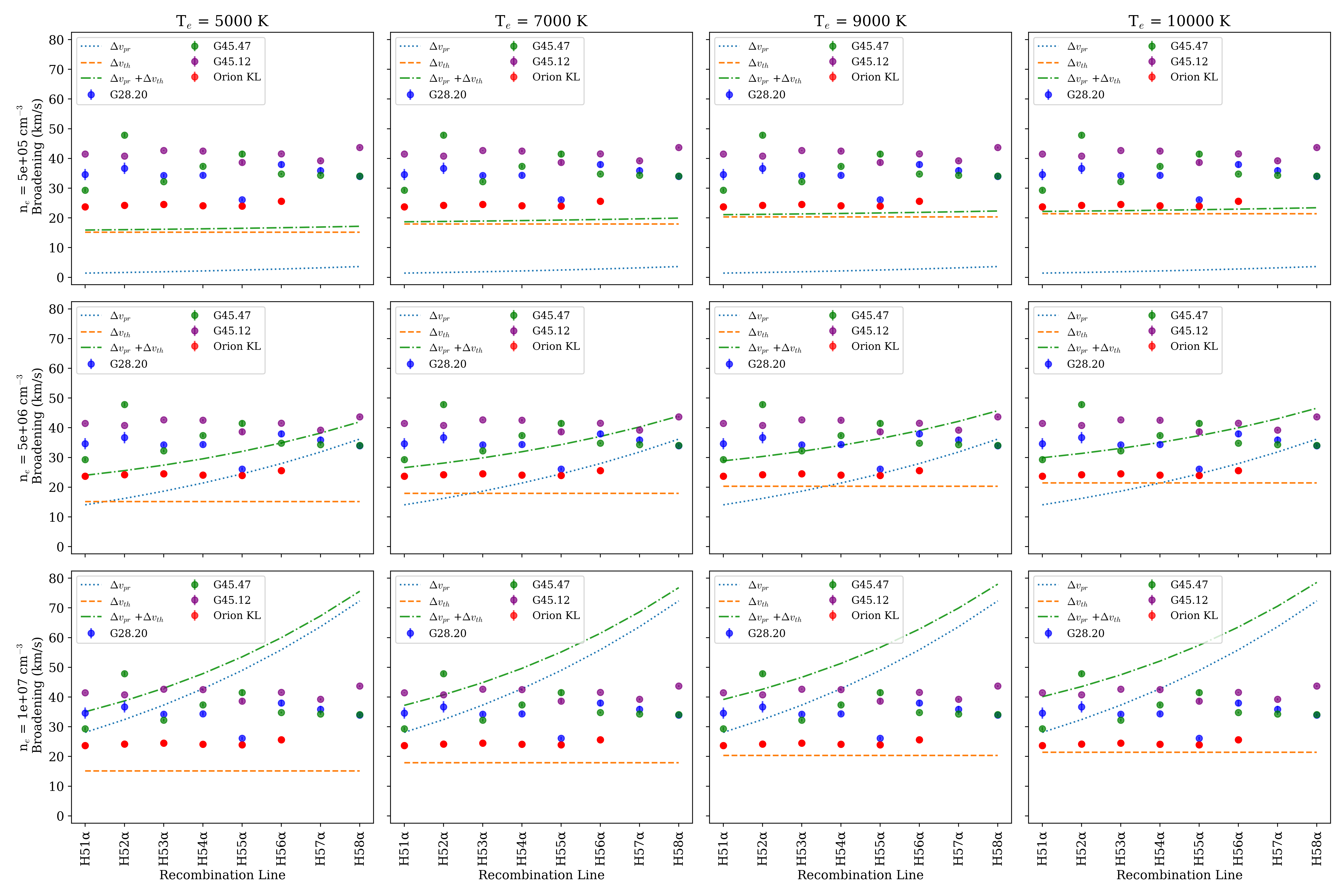}
\caption{Comparison between observed and theoretical line widths of H$\alpha$ lines for different electron densities and temperatures. In each panel, the blue dotted line represents the contribution from pressure broadening, the orange dashed line represents thermal broadening, and the green dash-dotted line depicts the combined contribution from thermal and pressure broadening. Line width data of Orion KL is taken from \citep{liu2022}.}
\label{fig:dens-temp-corr-halpha}
\end{figure*}

\begin{figure*}
\centering
\includegraphics[width=\textwidth]{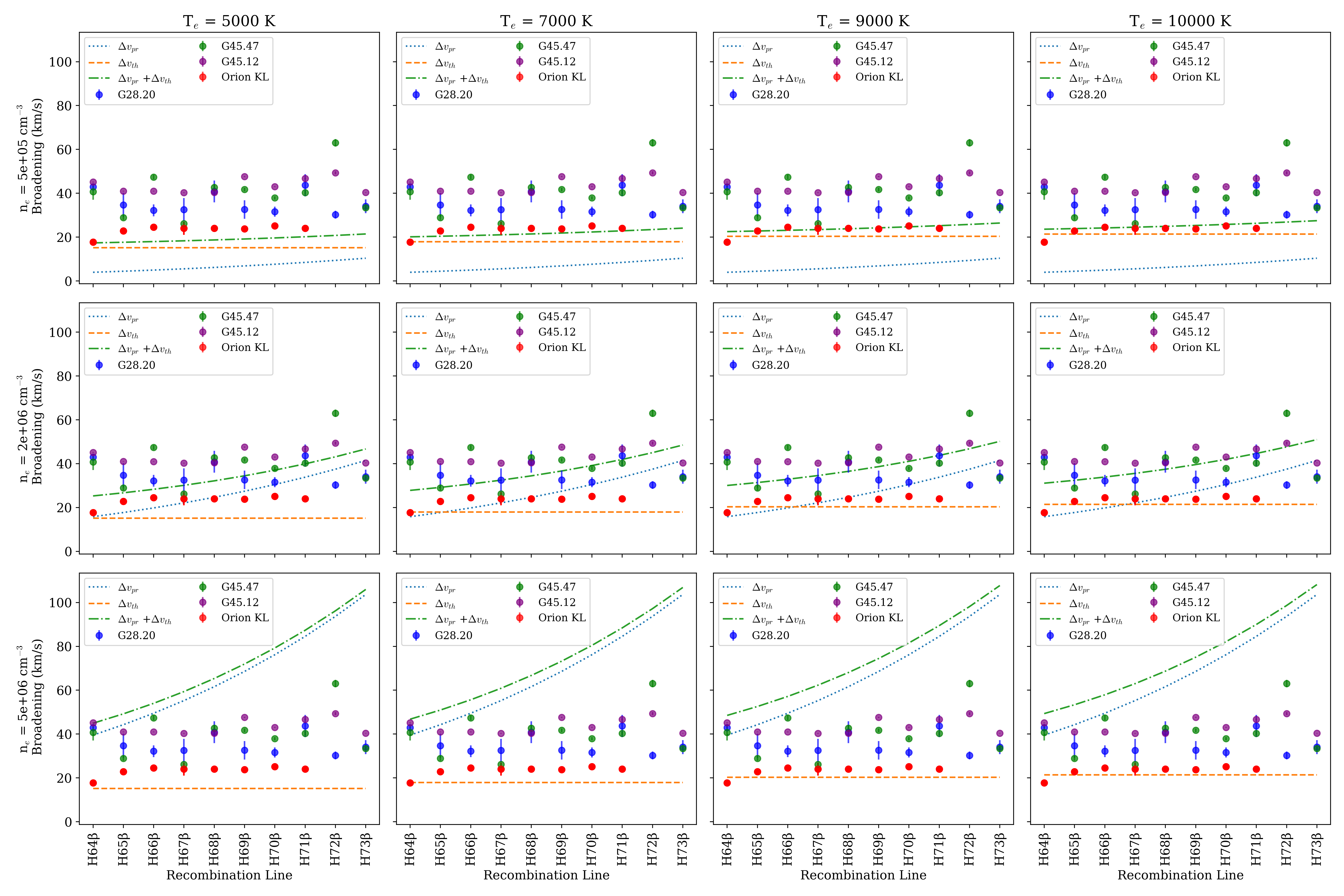}
\caption{Comparison between observed and theoretical line widths of H$\beta$ lines for different electron density and temperature. The blue dotted line represents the contribution from pressure broadening, the orange dashed line represents the thermal broadening, and the green dash-dotted line depicts the combined contribution from thermal and pressure broadening. Line width data of Orion KL is taken from \citep{liu2022}.}
\label{fig:dens-temp-corr-hbeta}
\end{figure*}

The natural broadening due to quantum uncertainty $\Delta E$ of the energy level is negligible for radio and (sub)mm RLs. The thermal distributions of velocities of both small pockets and individual particles of gas (i.e., ``microturbulence'') produce a Gaussian contribution to the broadening. Neglecting microturbulence, the thermal line width is  
\begin{equation}
\Delta v_\mathrm{th} = \biggl ( 8 \ln 2 \frac{k_B T_e}{m_\mathrm{H}} \biggr )^{1/2},  
\end{equation}
where $k_B$ is the Boltzmann constant and $m_\mathrm{H}$ is the mass of the hydrogen atom. Here, all the gas is assumed to be thermalized to the electron temperature $T_e$. The prescription of thermal line width is independent of electron density and proportional to the square root of the temperature. Thermal line widths (orange dashed lines) for different H$\alpha$ and  H$\beta$ transitions with varying electron temperatures are shown in Fig.~\ref{fig:dens-temp-corr-halpha} (also see Fig. \ref{fig:dens-temp-corr-hbeta} for H$\beta$ transitions). The figure shows the thermal line widths for three sets of electron temperatures and electron densities. For instance, assuming $T_e=7,000$ K, we obtain $\Delta v_\mathrm{th}=18.54$ km s$^{-1}$.

Pressure broadening shows a Lorentzian shape, increasing with density and quantum number. Collisions with ions and electrons contribute differently to the broadening, with the ion contribution taking the form following \citet{madrid2012}

\begin{equation}
\Delta v_\mathrm{pr,i} \approx \biggl (N_i \frac{c}{\nu_0}  \biggr ) 
(0.06 + 2.5\times10^{-4}T_e) \biggl ( \frac{n+1}{100} \biggr )^{\gamma_i} 
\biggl ( 1 + \frac{2.8 \Delta n}{n+1} \biggr ), 
\end{equation}
where $\gamma_i = 6-2.7\times10^{-5} T_e -0.13(n+1)/100$. For $T_e=9000\:$K 
and $N_i=10^7$ cm$^{-3}$, the H$30\alpha$ line is virtually free from ion 
broadening (0.04 km s$^{-1}$, and decreases close to linearly with decreasing 
$T_e$), whereas the H$53\alpha$ line is broadened by 5.1 km s$^{-1}$. Under most conditions, collisions with electrons dominate over those with ions, resulting in a broadening with a width given by \citet{madrid2012}
\begin{equation}
\Delta v_\mathrm{pr,e} \approx \biggl ( 8.2N_e \frac{c}{\nu_0}  \biggr ) 
 \biggl ( \frac{n+1}{100} \biggr )^{4.5} 
\biggl ( 1 + \frac{2.25 \Delta n}{n+1} \biggr ).  
\end{equation}
For $n_e=10^7\:$cm$^{-3}$, the electron broadening for the H53$\alpha$ and 
H$30\alpha$ are $\approx 37.3$ km s$^{-1}$  and 0.6 km s$^{-1}$, respectively. Figure \ref{fig:dens-temp-corr-halpha} shows the pressure broadening effect for different electron temperatures and densities. It shows the increasing trend of line width with both electron density and quantum number of the transitions. 

The final source of broadening is due to bulk motions of the ionized gas ($\Delta v_\mathrm{dyn}$), which may arise from outflows or winds, infall or accretion, and rotational dynamics. Hence, the total RL width ($\Delta v$) has contributions from thermal broadening ($\Delta v_\mathrm{th}$), dynamical broadening from macroscopic gas motions  ($\Delta v_\mathrm{dyn}$), and pressure broadening ($\Delta v_\mathrm{pr}$). The FWHM of the combined line profile is given by \cite{Madrid2009}
\begin{equation}
\Delta v \approx  0.534\Delta v_\mathrm{pr} +(\Delta v_\mathrm{dyn}^2 + 
\Delta v_\mathrm{th}^2 + 0.217\Delta v_\mathrm{pr}^2)^{1/2}. 
\end{equation}

The variation of thermal and pressure broadening with different electron temperatures and densities is shown in Figs. \ref{fig:dens-temp-corr-halpha} and \ref{fig:dens-temp-corr-hbeta}. These figures compare the measured line widths and theoretical estimates of broadening for H$\alpha$ and H$\beta$ lines. From these comparisons, we can examine the impact of various choices of electron temperature and density. The top and bottom panels of either figure show that with low ($\rm{\sim 1\times10^{5}}\ cm^{-3}$) or high ($\rm{\sim 1\times10^{7}}\ cm^{-3}$) electron density, the theoretical prediction, such as the combined thermal and pressure broadening effect, cannot explain the observed line widths for sources G45.12, G45.47, and G28.20, even with higher electron temperatures (10,000 K). For the low electron density case, it is underpredicted, and for high electron density, it is overpredicted. However, we find reasonably good matches between the theoretical estimation and observations at intermediate electron densities ($\rm{\sim 1-5\times10^{6}}\ cm^{-3}$) and electron temperatures of around 8,000-10,000 K. On the other hand, it is clear from Figs. \ref{fig:dens-temp-corr-halpha} and \ref{fig:dens-temp-corr-hbeta} that the pressure broadening shows an increasing trend of line width with quantum number n and electron density, whereas our observed line width for different recombination lines is more or less similar, which indicates that dynamical broadening would be important in this scenario. We did not include the dynamical broadening in the comparison because we do not know the microscopic motion of the gas. Nevertheless, we note that the level of dynamical broadening is degenerate with contributions from thermal broadening. Thus, low-temperature and low-density models could yield acceptable fits to the data when a dynamical broadening component is included (see Sec. \ref{sec:4}). High-resolution data of the HRLs are needed to search for potential spatially resolved velocity gradients that may yield more direct constraints on the level of dynamical broadening.

\subsection{Electron temperature and density \label{sec:3.4}}
Since our observations are single-pointing, single-dish measurements, we do not have accompanying continuum data. As a result, it is challenging to determine the electron temperature directly from line-to-continuum ratios. Therefore, we primarily relied on an empirical prescription based on the galactocentric distance to estimate the electron temperature for the different sources. Due to the gradient of metallicity with Galactocentric distance ($d_G$), the electron temperature in HII regions is known to vary. This variation has been parameterized as
\citep{Zhang2023}:
\begin{equation}
  T_e(\rm {K}) = (6271\pm481) + (135\pm83) (d_G / {\rm kpc}) .  
\end{equation}
The Galactocentric distances of each source are provided in Table \ref{tab:sources}.
For G28.20 ($D_G = 4.10$ kpc), the obtained electron temperature ($T_e$) is $6825 \pm 589$ K. Similarly, electron temperatures are estimated at $7054 \pm 681$ K and  $7050 \pm 679$ K for G45.47 ($D_G = 5.80$ kpc) and G45.12 ($D_G = 5.77$ kpc) , respectively. 

The electron density may be estimated following the equation given by \cite{keto08a}:
\begin{equation}
    \frac{\Delta\nu_L}{\Delta\nu_t}=\frac{1.2}{\Delta n}\ \ \Bigg(\frac{n_e}{10^5\:{\rm cm}^{-3}}\Bigg)\ \Bigg(\frac{n}{92}\Bigg)^7,
\end{equation}
where $\Delta\nu_L$ is the Lorentzian width and $\Delta\nu_{\rm th}$ is the thermal width, and $\Delta n = 1$ for H$\alpha$ transitions and $\Delta n = 2$ for H$\beta$ transitions. Using the above prescription, we measured the electron density of all transitions, which is summarized in Table \ref{tab:electron-dens}. The measured electron densities for all sources are $\sim 10^6\:{\rm cm}^{-3}$. 

\begin{table}[hbt]
\centering
\caption{Calculated electron densities ($n_e$) for different transitions across all sources.}
\label{tab:ne_all_sources}
\begin{tabular}{cccl}
\hline
Transition & G28.20 & G45.12 & G45.47 \\
 & (n$_e$ cm$^{-3}$) & (n$_e$ cm$^{-3}$) & (n$_e$ cm$^{-3}$) \\

\hline
H51$\alpha$ & 1.01e+07 & 1.19e+07 & 8.44e+06 \\
H52$\alpha$ & 9.38e+06 & 1.03e+07 & 1.20e+07 \\
H53$\alpha$ & 7.66e+06 & 9.39e+06 & 7.09e+06 \\
H54$\alpha$ & 6.74e+06 & 8.21e+06 & 7.22e+06 \\
H56$\alpha$ & 5.78e+06 & 6.22e+06 & 5.21e+06 \\
H57$\alpha$ & 4.82e+06 & 5.19e+06 & 4.54e+06 \\
H58$\alpha$ & 4.04e+06 & 5.12e+06 & 3.99e+06 \\
H64$\beta$  &    --     & 5.30e+06 & 4.79e+06 \\
H65$\beta$  & 3.72e+06  & 4.32e+06 & 3.05e+06 \\
H66$\beta$  & 3.10e+06  & 3.88e+06 & 4.49e+06 \\
H67$\beta$  & 2.82e+06 & 3.44e+06 & 2.23e+06 \\
H68$\beta$  & 3.20e+06 & 3.11e+06 & 3.28e+06 \\
H69$\beta$  & 2.30e+06 & 3.31e+06 & 2.89e+06 \\
H70$\beta$  & 2.02e+06 & 2.70e+06 & 2.38e+06 \\
H71$\beta$  & 2.52e+06 & 2.66e+06 & 2.29e+06 \\
H72$\beta$  & 1.58e+06 & 2.54e+06 & 3.25e+06 \\
H73$\beta$  & 1.62e+06 & 1.89e+06 & 1.56e+06 \\
\hline
\end{tabular}
\label{tab:electron-dens}
\end{table}

\begin{table*}[hbt]
\scriptsize
\centering
\caption{Thermal and dynamical broadening of hydrogen recombination lines. \label{tab:recomb_broadening_combined}}
\begin{tabular}{|c|p{1.0cm}|
cc|cc|cc|cc|}
\hline
Transition & Thermal Width (km/s) 
& \multicolumn{2}{c|}{G45.12+0.13} 
& \multicolumn{2}{c|}{G45.47+0.05} 
& \multicolumn{2}{c|}{G28.20--0.05} 
& \multicolumn{2}{c|}{Orion KL} \\
\cline{3-10}
& & Measured FWHM & Dyn. Width
& Measured FWHM & Dyn. Width  
& Measured FWHM & Dyn. Width  
&Measured FWHM & Dyn. Width \\
& & \multicolumn{2}{|c|}{(km/s)} & \multicolumn{2}{|c|}{(km/s)} & \multicolumn{2}{|c|}{(km/s)} & \multicolumn{2}{|c|}{(km/s)} \\
\hline
H51$\alpha$ & 18.54 & 41.46 & 22.92 & 29.30 & 10.76 & 34.60 & 16.06 & 23.70 & 5.16 \\
H52$\alpha$ & 18.54 & 40.78 & 22.24 & 47.83 & 29.29 & 36.68 & 18.14 & 24.20 & 5.66 \\
H53$\alpha$ & 18.54 & 42.66 & 24.12 & 32.21 & 13.67 & 34.24 & 15.70 & 24.50 & 5.96 \\
H54$\alpha$ & 18.54 & 42.51 & 23.97 & 37.37 & 18.83 & 34.36 & 15.82 & 24.09 & 5.55 \\
H55$\alpha$ & 18.54 & 38.65 & 20.11 & 41.48 & 22.94 & 26.09 & 7.55 & 23.94 & 5.40 \\
H56$\alpha$ & 18.54 & 41.55 & 23.01 & 34.78 & 16.24 & 37.97 & 19.43 & 25.60 & 7.06 \\
H57$\alpha$ & 18.54 & 39.22 & 20.68 & 34.30 & 15.76 & 35.88 & 17.34 & -- & -- \\
H58$\alpha$ & 18.54 & 43.68 & 25.14 & 34.07 & 15.53 & 33.93 & 15.39 & -- & -- \\
H64$\beta$  & 18.54 & 45.10 & 26.56 & 40.74 & 22.20 & -- & -- & 17.70 & 0.00 \\
H65$\beta$  & 18.54 & 40.97 & 22.43 & 28.90 & 10.36 & 34.68 & 16.14 & 22.80 & 4.26 \\
H66$\beta$  & 18.54 & 40.96 & 22.42 & 47.38 & 28.84 & 32.17 & 13.63 & 24.50 & 5.96 \\
H67$\beta$  & 18.54 & 40.26 & 21.72 & 26.17 & 7.63  & 32.51 & 13.97 & 24.00 & 5.46 \\
H68$\beta$  & 18.54 & 40.38 & 21.84 & 42.68 & 24.14 & 40.87 & 22.33 & 24.00 & 5.46 \\
H69$\beta$  & 18.54 & 47.59 & 29.05 & 41.69 & 23.15 & 32.60 & 14.06 & 23.80 & 5.26 \\
H70$\beta$  & 18.54 & 43.02 & 24.48 & 37.86 & 19.32 & 31.58 & 13.04 & 25.10 & 6.56 \\
H71$\beta$  & 18.54 & 46.71 & 28.17 & 40.26 & 21.72 & 43.66 & 25.12 & 24.00 & 5.46 \\
H72$\beta$  & 18.54 & 49.32 & 30.78 & -- & -- & 30.24 & 11.70 & -- & -- \\
H73$\beta$  & 18.54 & 40.33 & 21.79 & 33.40 & 14.86 & 34.06 & 15.52 & -- & -- \\
\hline
\end{tabular}
Note- Thermal line widths are estimated using a fixed electron temperature of 7500 K, which was estimated based on their Galactocentric distance and roughly similar values for all sources within uncertainties. For this, we assume a negligible effect from pressure broadening.
\label{tab:vth-vdy}
\end{table*}

\begin{figure*}
\centering
\includegraphics[width=\textwidth]{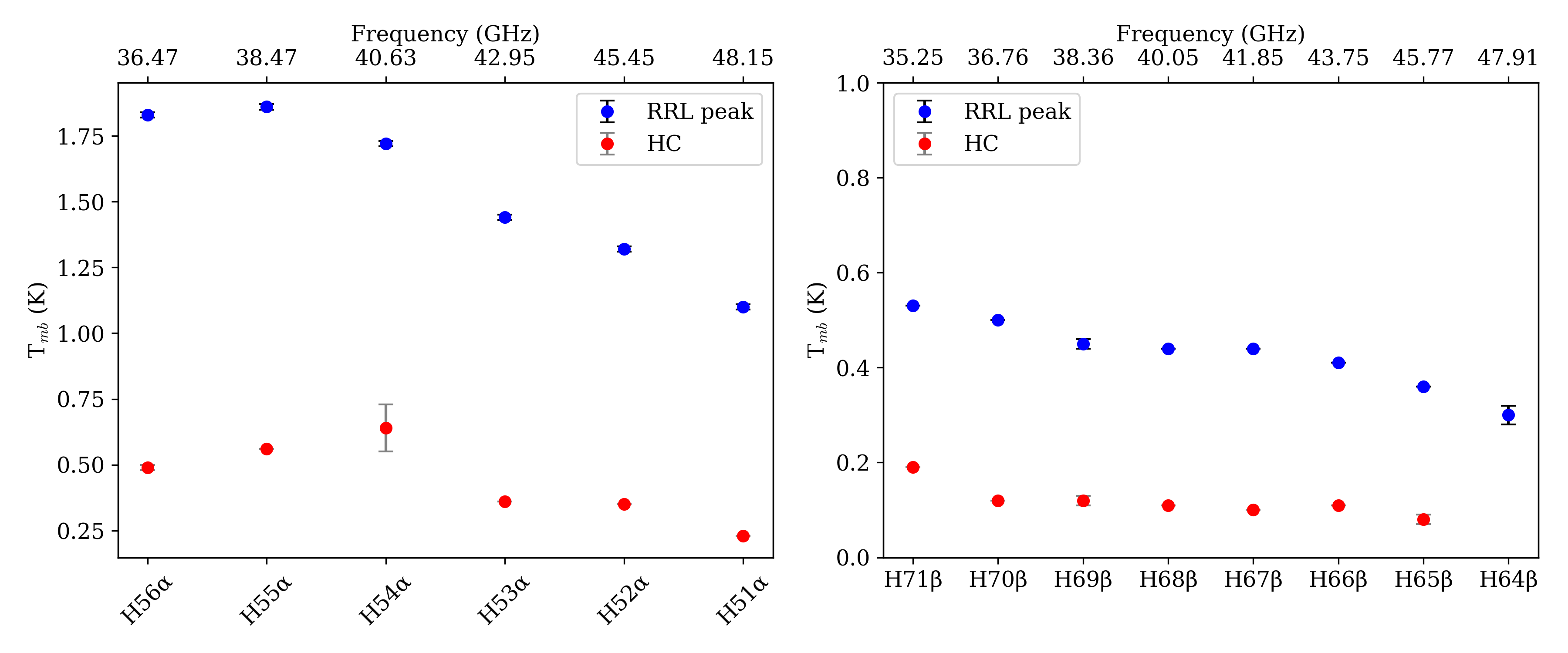}
\caption{Line intensities of the H$\alpha$ and H$\beta$ recombination lines based on ALMA data, measured toward the radio recombination line peak and the hot core region of Orion KL.}
\label{fig:line-int-alma}
\end{figure*}

\begin{figure}
\centering
\includegraphics[width=\linewidth]{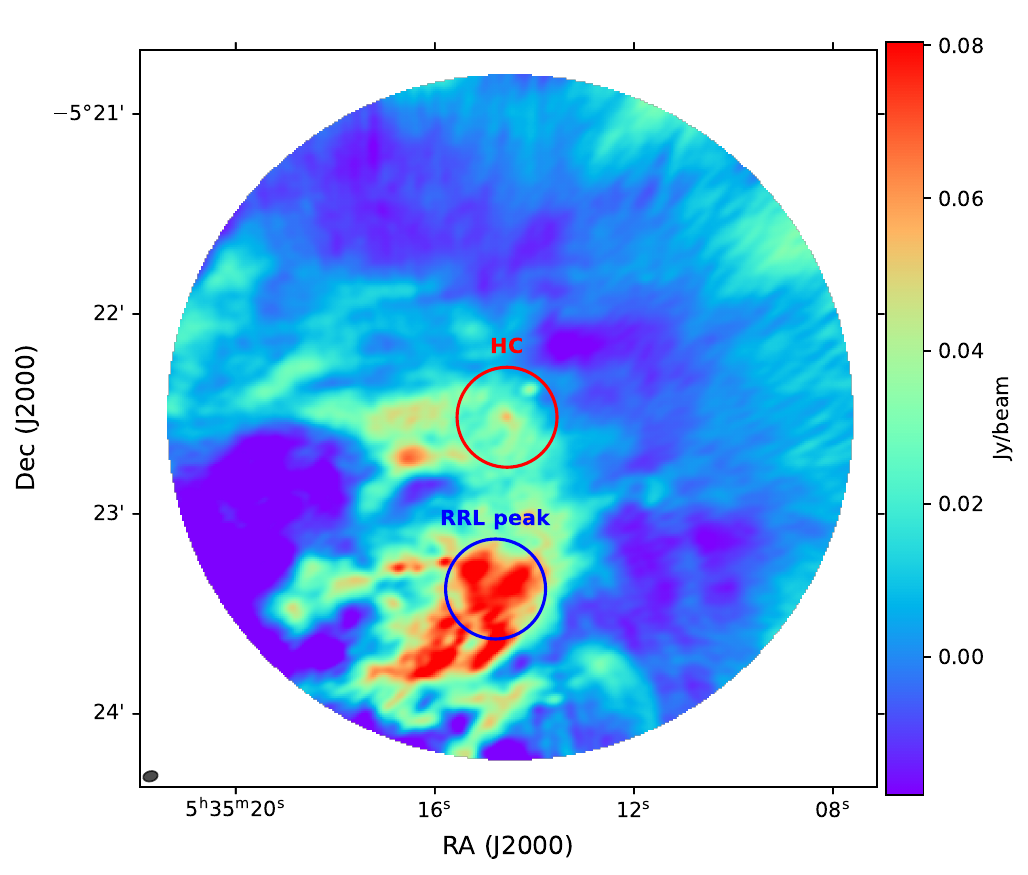}
\caption{7.1 mm continuum image of Orion KL. Blue and red circles indicate the 30$^{\prime\prime}$ diameter regions centered on the hot core and the RRL peak, from which spectra were extracted to measure the line intensity, as shown in Fig. \ref{fig:line-int-alma}}
\label{fig:cont-7mm}
\end{figure}

\section{Discussion\label{sec:4}}
The measured line widths of the H$\alpha$ and H$\beta$ transitions towards G28.20, G45.45, and G45.12 are $\sim$ 35-40 $\rm{km s^{-1}}$ (see Table \ref{tab:lines}). However, the measured line widths of these transitions in Orion KL are around $\sim$ 25 km/s \citep{liu2022}. These differences could be due to several factors. The electron temperature determines thermal broadening, which contributes to the overall line width of the spectrum. The estimated electron temperature in Orion KL is 7391 $\pm$ 840 K, considering a Galactocentric distance of 8.3 kpc \citep{milam2005}. For our target sources, we obtained similar temperatures (see \ref{sec:3.4}). However, we observe apparent differences in line width between Orion KL and other sources, despite their similar electron temperatures. Therefore, thermal broadening alone cannot explain these differences. These variations suggest different pressure-broadening effects caused by distinct electron density levels in the sources or different levels of dynamical broadening. Figures \ref{fig:dens-temp-corr-halpha} and \ref{fig:dens-temp-corr-hbeta} show a comparison between observed and theoretically measured line widths for H$\alpha$ and H$\beta$ transitions, respectively. These figures compare the theoretical and observed line widths for varying electron density and temperature. Thermal broadening is independent of the quantum number of RRLs and is proportional to the magnitude of the temperature for the same species. In contrast, for a low electron density (e.g., $\rm{\sim 5\times10^{5}\ cm^{-3}}$), pressure broadening has a minimal effect on the line width, whereas for a high electron density (e.g., $\rm{\sim 1\times10^{7}\ cm^{-3}}$), it has a significant impact. The pressure broadening highly depends on the electron density and increases with the quantum numbers of the transition. We find good agreement between the observed and theoretical line widths for the H$\alpha$ and H$\beta$ lines in G45.12, G45.47, and G28.20 with a high electron temperature $\sim$ 8000-10000 K and a moderate electron density $\rm{\sim1-5\times10^{6}\ cm^{-3}}$. 

On the other hand, our comparison (see Figs. \ref{fig:dens-temp-corr-halpha} and \ref{fig:dens-temp-corr-hbeta}) clearly shows a dependence on the quantum number of pressure broadening, which was not evident from the measured line widths, as they mostly show similar values for all HRLs (see Table \ref{tab:lines}). In this scenario, dynamical broadening would then play the crucial role. The photoionized disk-wind outflows around forming massive stars are important for understanding the line profile broadening of recombination lines \citep{2016ApJ...818...52T,Martinez2023,Martinez2024}. 

Furthermore, we compare our results with those of Orion KL. Here, the line widths of RRLs measured show only a minor effect from pressure broadening, since Orion KL has an electron density at least one order of magnitude lower than the sources reported in this study. The best-fit electron density and temperature for Orion KL are $\rm{1-5\times10^{5}\ cm^{-3}}$ and 8000–10000 K, respectively. The best fitted electron density is consistent with the inferred electron density ($\rm{\sim 1\times10^{6}\ cm^{-3}}$) obtained with the given prescription in Sec. \ref{sec:3.4} (also see Table \ref{tab:electron-dens}). Since we did not observe an increasing trend in the line widths of the detected HRLs, we assume that the widths are less affected by pressure broadening, indicating probably a low electron density ($\sim 10^{5}\ \mathrm{cm^{-3}}$). Hence, the measured line width is most likely due to the thermal and dynamical broadening. Therefore, we estimated the dynamical broadening for all sources by subtracting the thermal width from the measured line width. All values are summarized in Table \ref{tab:vth-vdy}.

Figure \ref{fig:int-variation} shows the intensity variation with frequency (recombination lines transitions in descending order of quantum numbers) for our target sources along with the existing result of Orion KL \citep{liu2022}. We found an increasing trend in line intensity with frequency for both H$\alpha$ and H$\beta$ lines in G45.12 and G28.20, with a slope of 1. For G45.47, the trend is relatively flat. The increasing intensity trend of the RRLs may result from inozied wind, partially optically thick emission, density, and temperature gradients, or beam dilution effects.  Weak non-LTE effects (e.g., stimulated emission) can enhance the line intensity at higher frequencies, particularly in dense and dynamic HII regions. The rising intensity likely reflects the compact and dense nature of these YSOs' HII regions, with higher frequencies tracing regions of elevated $n_e$ or $T_e$. Additionally, the increasing trend may partly result from improved resolution at higher frequencies (e.g., around 50\,GHz), where the synthesized beam is slightly better compared to the source size. The beam of the Yebes 40m telescope varies significantly across the Q band, ranging from 55$^{\prime\prime}$ at the lowest frequencies to 35$^{\prime\prime}$ at the highest. If the source is compact relative to the beam size, this variation can substantially affect the observed intensity of the HRLs. 

In contrast, the Orion KL exhibits the opposite trend compared to our target sources, where the intensity of HRLs decreases with frequency in the 30-50 GHz range. This behavior suggests either optical depth effects or beam dilution, as it varies with frequency. In single dish observations of Orion KL, the primary beam size was 30$^{\prime\prime}$ at 40 GHz \citep{liu2022}. To further investigate, we examined the trend in HRL intensities in Orion KL using ALMA Band 1 line survey data (see Fig. \ref{fig:line-int-alma}) and found a similar result to that obtained from single-dish observations. For this analysis, we selected two different locations (see Fig. \ref{fig:cont-7mm}): one near the hot core (HC) peak and another near the recombination line (RRL) peak, where HRLs are particularly strong. We measured the line intensities from spectra extracted within a 30$^{\prime\prime}$-diameter aperture. The opposite trend in the Orion KL can be attributed to specific local conditions and physical processes affecting the source. The local electron density and temperature can affect the population of energy levels differently. Interestingly, while lower-order transitions like H51$\alpha$ are expected to be more intense than higher-order ones (e.g., H56$\alpha$) under low-density conditions due to collisional excitation, we observe the opposite trend. The observed reversal trend suggests that non-LTE effects, higher electron densities, or optical depth effects may be influencing the excitation and radiative transfer of the HRLs in this region. The decreasing intensity trend with frequency may be due to a more extended, lower-density, or more evolved HII region. The narrower line widths of RRLs towards Orion KL compared to our target sources suggest less dynamical activity. This region may be older or less massive, with a lower-density HII region, which explains both the lower intensity and narrower lines. Dust absorption could also contribute to the reduced intensity, especially if a dense dust envelope surrounds the source. Additionally, the decreasing trend might be influenced by stimulated emission in higher-$n$ RRLs (i.e., those at lower frequencies), as these effects are more pronounced in transitions with high principal quantum numbers and lower-density environments. Nonetheless, High-spatial-resolution observations and modeling are necessary to understand the specific reasons for this behavior, and such studies can provide valuable insights into the diverse nature of ionized regions.

\section{Conclusions \label{sec:5}}

We have conducted a comprehensive spectral line survey targeting six high-mass protostars, employing the Q-band receiver of the Yebes 40m telescope. This survey spanned a frequency range of 31.5 to 50 GHz and marks the first Q-band line survey conducted toward three high-mass sources: G45.12, G28.20, and G45.47. The key findings from this study are outlined below.

\begin{itemize}
\item We detected 18 recombination lines using the RT40m Q-band observations toward three high-mass protostars. Among these lines, eight correspond to H$\alpha$ ($n =$ 51 to 58) transitions, and the remaining ten are H$\beta$ ($n =$ 64 to 73) transitions.  We determine the line parameters, including line width, peak line intensity, and integrated intensity. 

\item We obtained an electron density of approximately $\rm{1-5\times10^6\ cm^{-3}}$ and a temperature of around 8000–10000 K for G45.12, G28.20, and G45.47. In contrast, the electron density for Orion KL is about an order of magnitude lower, at $\rm{\sim 1-5\times10^5\ cm^{-3}}$, while the temperature remains in a similar range of $\sim$8000–10000 K. However, we did not find the line trend that broadens with the quantum number, as shown in Figs. \ref{fig:dens-temp-corr-halpha} and \ref{fig:dens-temp-corr-hbeta}, which is prominent for the high electron density $\ge$ $\rm{\sim 10^5\ cm^{-3}}$. Therefore, pressure broadening might have a minor effect compared to thermal and dynamical broadening. Hence, our constrained electron density can be considered as an upper limit. 

\item The observed line widths cannot be accounted for by thermal broadening alone. We consider that dynamical broadening plays a significant role in explaining the width of the HRLs, suggesting the presence of turbulence, rotation, or outflow broadened line widths of ionized gas. The potential contribution of dynamical broadening for all sources has been estimated by subtracting the thermal line width from the measured line widths.

\item The intensity profiles of the H$\alpha$ and H$\beta$ transitions show an increasing trend with frequency for G45.12 and G28.20, with slope $\sim$ 1 obtained from a power-law fit, for G45.47 it is almost flat. In contrast, Orion KL exhibits a decreasing intensity trend with increasing frequency. This could be due to several reasons, such as differences in physical structures (e.g., electron density), optically thin or thin lines, stimulated emission, maser effects, and beam dilutions. Further investigation is required with the systematic high-spatial resolution data.

\end{itemize}    
    
\begin{acknowledgements}
Based on observations carried out with the Yebes 40m telescope (projects 21A-004 and 23A017). The 40m radiotelescope at Yebes Observatory is operated by the Spanish Geographic Institute (IGN, Ministerio de Transportes, Movilidad y Agenda Urbana). This paper makes use of the following ALMA data: ADS/JAO.ALMA\#2011.0.00022.SV. ALMA is a partnership of ESO (representing its member states), NSF (USA) and NINS (Japan), together with NRC (Canada), NSTC and ASIAA (Taiwan), and KASI (Republic of Korea), in cooperation with the Republic of Chile. The Joint ALMA Observatory is operated by ESO, AUI/NRAO and NAOJ. P.G. and M.S. acknowledge the ESGC project (project No. 335497) funded by the Research Council of Norway. K.T. is supported by JSPS KAKENHI grant Nos. 21H01142, 24K17096, and 24H00252. M.G.-G. acknowledges support from the grant PID2023-146056NB-C21 (CRISPNESS) funded by MICIU/AEI/10.13039/501100011033 and by ERDF/EU. I.J-.S acknowledges funding from grant PID2022-136814NB-I00 funded by the Spanish Ministry of Science, Innovation and Universities/State Agency of Research MICIU/AEI/ 10.13039/501100011033 and by “ERDF/EU”. R.F. acknowledges financial support from the Severo Ochoa grant CEX2021-001131-S MICIU/AEI/ 10.13039/501100011033 and PID2023-146295NB-I00. C-Y.L. acknowledges the financial support through the INAF Large Grant The role of MAGnetic fields in MAssive star formation (MAGMA). G.B. acknowledges support from the PID2020-117710GB-I00 grant funded by MCIN/AEI/10.13039/501100011033 and from the PID2023-146675NB-I00 (MCI-AEI-FEDER, UE) program. BALG is supported by the German Research Foundation (DFG) in the form of an Emmy Noether Research Group - DFG project \#542802847 (GA 3170/3-1). 
\end{acknowledgements}

\bibliographystyle{aa}
\bibliography{References}{}

\end{document}